\documentclass[11pt,twoside]{article}

\usepackage{asp2004}
\usepackage{epsf}
\usepackage{psfig}
\usepackage{lscape}
\usepackage{graphicx}

\begin{document}
\title{{\em Spitzer/MIPS} $24 \,\mu \rm m$ galaxies: the link to near-IR galaxies and the cosmic IR background}   
\author{K.I. Caputi, H. Dole, G. Lagache, J.-L. Puget}   
\affil{Institut d'Astrophysique Spatiale (IAS), b\^at. 121, F-91405 Orsay, France; 
	   Universit\'e Paris-Sud 11 and CNRS (UMR 8617)}    

\begin{abstract} 
We present the results of our most recent works on {\em Spitzer/MIPS} $24 \,\mu \rm m$ galaxies. Through a multiwavelength analysis, we study different properties (redshifts, luminosities, stellar masses) characterising the sources which produce the bulk of the mid-IR background. From a comparative study with the total population of $K_s$-selected galaxies, we determine that $24 \,\mu \rm m$ sources account for an important fraction of the most massive galaxies present at different redshifts. On the other hand, we determine that $24 \,\mu \rm m$ galaxies also produce most of the energy contained in the far-IR cosmic background at 70 and $160 \,\mu \rm m$.  Furthermore, we are able to set tight constraints on the Cosmic Infrared Background (CIB) spectral energy distribution (SED). Our results help to clarify the links between these presumably different IR galaxy populations. 
\end{abstract}


\section{Introduction}

The Cosmic Infrared Background (CIB) accounts for roughly half of the total energy produced by extragalactic sources (e.g. Hauser \& Dwek 2001). Since the discovery  of the CIB (Puget et al. 1996), it has been recognised that the study of infrared (IR) extragalactic sources is fundamental to understand galaxy formation and evolution.  As IR emission is produced by the dust re-processing of UV/optical light,  IR sources constitute directly the signposts of star formation or accretion activity in the Universe.

 The determination of the properties characterising IR galaxies and the link between the IR and other galaxy populations have been a matter of study since the first IR missions ({\em IRAS, ISO}). With the advent of {\em Spitzer} (Werner et al. 2004), our comprehension of the nature and composition of the CIB has very much improved. Numerous works presented during this conference have shown how much progress we have made in 
understanding IR galaxies up to high redshifts $z \sim 4-5$.

In this paper, we summarize our recent studies of the properties of mid-IR galaxies and their contribution to other wavelength domains. In particular, in Section 3, we analyze the role of the most luminous mid-IR galaxies in the  evolution of the most massive $K_s$-band galaxies. In Section 4, we constrain the contribution of mid-IR galaxies to the far-IR background and give new estimates for the CIB. Finally, in Section 5, we briefly discuss the implications of our work.

\section{Properties of the sources composing the $24 \,\mu \rm m$  background}
\label{secprop}

Caputi et al.~(2006a) studied different properties of the sources composing the bulk of the $24 \, \rm \mu m$ background in the Great Observatories Origins Deep Survey / Chandra Deep Field South (GOODS/CDFS). Observations of the CDFS have been carried out with the Multiband Imaging Photometer
   for {\em Spitzer} (MIPS; Rieke et al.~2004), as part of the Guaranteed Time Observers (GTO) program. Using a deep ($K_s<21.5$, Vega) galaxy sample,  Caputi et al.~(2006a) identified $\sim 94\%$ of the sources  with  $S_\nu(24 \, \rm \mu m)>83 \, \rm \mu Jy$ in the GOODS/CDFS. Taking advantage of the excellent quality multiwavelength photometry and spectroscopic coverage of this field, they determined  the redshift distribution, IR luminosities and stellar masses characterising  $24 \, \rm \mu m$ galaxies.

\begin{figure}
\centering
\includegraphics[width=8cm]{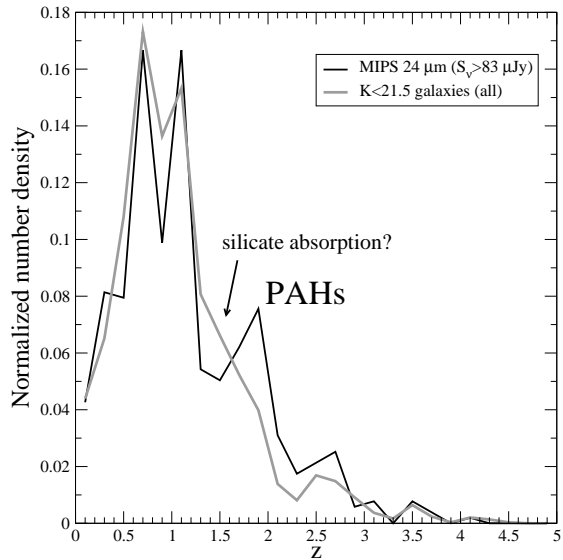}
\caption{ The normalized redshift distributions of the $S_\nu >83 \, \mu {\rm Jy}$  $\rm 24\, \mu m $ galaxies (black line), compared to the normalized redshift distribution of the total $K_s<21.5$ sample in the same field (grey line).  Figure taken from Caputi et al. (2006a).}
\label{fig1} 
\end{figure}

Figure~\ref{fig1} shows the normalized redshift distributions of the $S_\nu(\rm 24 \, \mu m) > 83 \, \rm \mu Jy$ galaxies in comparison to the distribution of all the $K_s$-band galaxies in the GOODS/CDFS. These populations are composed by $\sim 500$ and 3000 objects, respectively,  over an area of $\sim$130 arcmin$^2$.  Several features are present in both distributions, which are the consequence of known large-scale structure in the GOODS/CDFS. In contrast, we observe the existence of a bump in the redshift distribution of $24 \, \rm \mu m$ galaxies at redshift $z \sim 1.9$, which does not appear for the total $K_s<21.5$ galaxy population. This peak in the $24 \, \rm \mu m$ redshift distribution has been predicted by Lagache et al.~(2004) and is the consequence of the selection effect produced by the  presence of  PAH emission features entering the observed $\rm 24\,\mu m$ band.  Given the width of the $\rm 24\,\mu m$ filter (whose transmission covers the wavelength range $\sim 20-28 \, \mu \rm m$), both the 7.7 and the $8.6 \mu \rm m$ PAH lines could contribute to the redshift distribution peak observed at $z \sim 1.9$.  Our results show that  PAH molecules must already be present in a significant amount of star-forming galaxies at high redshifts.

\begin{figure}
\centering 
\includegraphics[width=14cm]{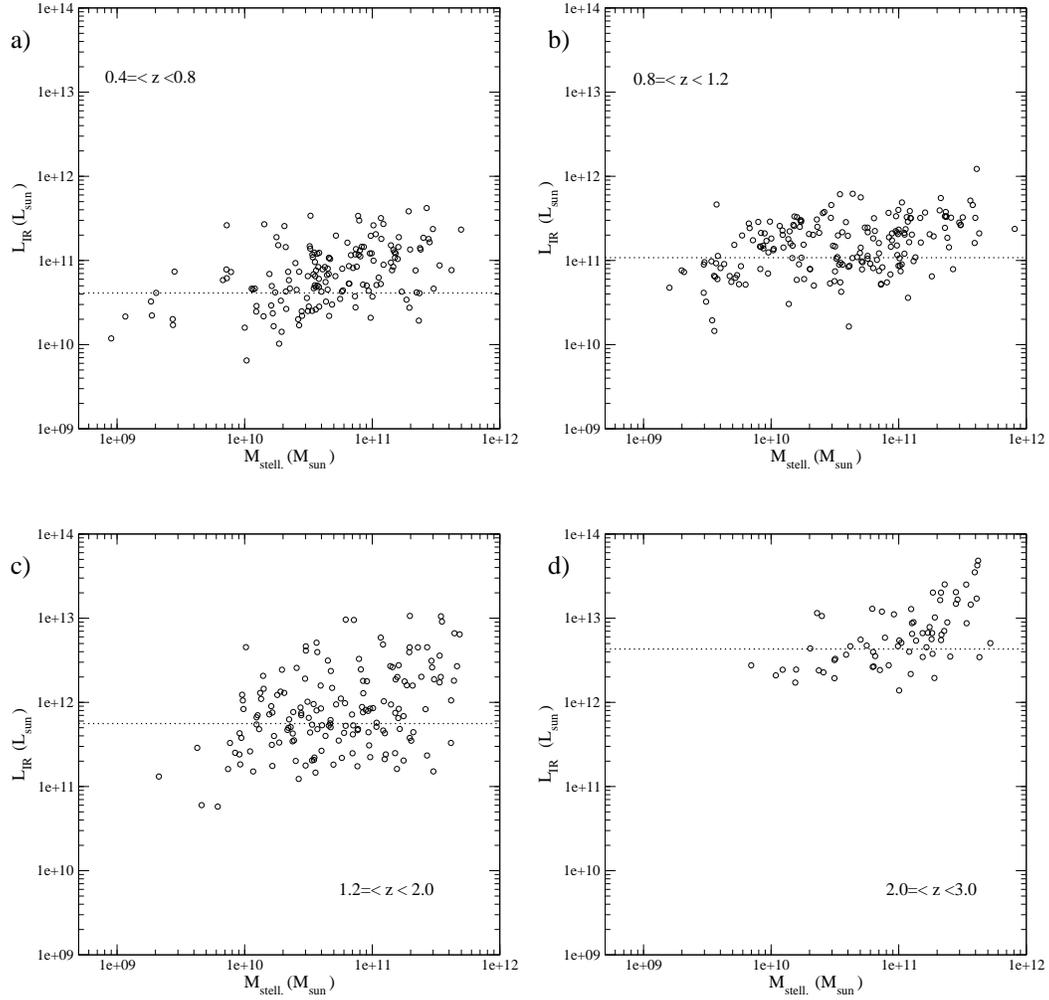}
\caption{The IR luminosity ($L_{IR}$) versus stellar mass ($M_{stell.}$) relation for $\rm 24\,\mu m$ galaxies,  at different redshifts. In each panel, the dotted line indicates the luminosity completeness  imposed by the $S_\nu = 83 \, \rm \mu Jy$ flux completeness limit of the $\rm 24\,\mu m$ sample.  Figure taken from Caputi et al.~(2006a).}
\label{fig2}
\end{figure}

Using  the empirical calibrations obtained by Chary \& Elbaz (2001) and Elbaz et al. (2002) between mid-IR luminosities and bolometric IR luminosities $L_{IR}=L(8-1000 \mu{\rm m})$, Caputi et al. (2006a) computed bolometric IR luminosity estimates for all their star-forming $\rm 24 \, \mu m$ galaxies. On the other hand, they modelled the optical/near-IR spectral energy distributions (SEDs) of these galaxies, including {\em Spitzer/IRAC (Infrared Array Camera)} $3.6$ and $4.5 \, \rm \mu m$ data. This allowed them to obtain rest-frame $K_s$-band luminosities and derived stellar masses.

Figure~\ref{fig2} shows the evolution of the IR luminosity ($L_{IR}$) versus stellar mass ($M_{stell.}$) plane as a function of redshift. Panel a) shows that most of the  $\rm 24 \, \mu m$ galaxies at redshifts $0.4 \leq z < 0.8$ have infrared luminosities $L_{IR}<10^{11}\, L_{\odot}$.  The maximum observed infrared luminosities increase with redshift, and luminous infrared galaxies (LIRGs) characterized by  $10^{11}\, L_{\odot}<L_{IR}<10^{12}\, L_{\odot}$ are the dominant  $\rm 24 \, \mu m$  population at redshifts  $0.8 \leq z < 1.2$ (cf. also Le Floc'h et al. 2005). The majority of the mid-IR sources at  $0.4 \leq z < 1.2$ are hosted by intermediate-mass galaxies with stellar masses $10^{10} \, M_{\odot} < M < 10^{11}  \, M_{\odot}$, although some more massive galaxies could also be classified as LIRGs at these redshifts. Within our surveyed area, there is virtually no ultra-luminous infrared galaxy (ULIRG) with $L_{IR}>10^{12}\, L_{\odot}$ at $z<1.2$. ULIRGs might be present at these low redshifts, but are indeed very rare (e.g. Flores et al.~1999).  At $1.2 \leq z \leq 2.0$, ULIRGs start to be the dominant population ($\sim$ 65\% at $S_\nu > 83 \mu {\rm Jy}$). Most of them are intermediate to high-mass galaxies.  At $z>2$, we observe sources with extremely high infrared luminosities  $10^{12}\, L_{\odot}<L_{IR}<10^{14}\, L_{\odot}$ mainly harboured by galaxies with stellar masses  $M > 10^{11} \, M_{\odot}$.

For star-forming galaxies, an IR luminosity $L_{IR} \sim 10^{11}\, (10^{12})\, L_\odot$  corresponds to a star-formation rate $SFR \sim  20 \, (200) \, M_\odot \, \rm yr^{-1}$ (Kennicutt 1998). The analysis of X-ray data and IRAC colour-colour diagrams suggests that only a minor fraction of IR galaxies could be mainly driven by quasar activity (Caputi et al. 2006b). Thus, the bulk of the IR emission in most $24 \, \rm \mu m$ galaxies must be produced by star-formation activity, whose rates can achieve extremely high values ($SFR>500-1000 \, M_\odot \rm yr^{-1}$) at high redshifts.  

The large $SFR$ characterising some systems at $z \sim 2-3$ imply that a few starburst episodes lasting $10^7-10^8$ yr might be sufficient to build up the stellar mass of some massive ($\sim 10^{11}  \, M_{\odot}$)  galaxies at these redshifts. Thus, the burst-like mode of star formation appears to have been a very efficient way of constructing  massive galaxies in the past.  On the contrary, at lower redshifts, the efficiency of the burst-like mode of star-formation to construct entire galaxies is limited to lower mass systems. A significant fraction ($\sim 30\%$)  of massive galaxies experience star-formation activity at lower redshifts, but this star formation only produces an additional minor amount of the stellar mass already present in these systems (Caputi et al.~2006a,b). 

\section{The role of $24 \, \rm \mu m$ sources in the evolution of $K_s$-band galaxies}
\label{secks}

Near-IR surveys are traditionally used to trace stellar mass at different redshifts.  The study of the
role of mid-IR galaxies within the context of the total near-IR-selected galaxy population should allow, then,  to understand the importance of star formation and accretion activity in galaxies of different stellar mass.

\begin{figure}
\centering
\includegraphics[width=14cm]{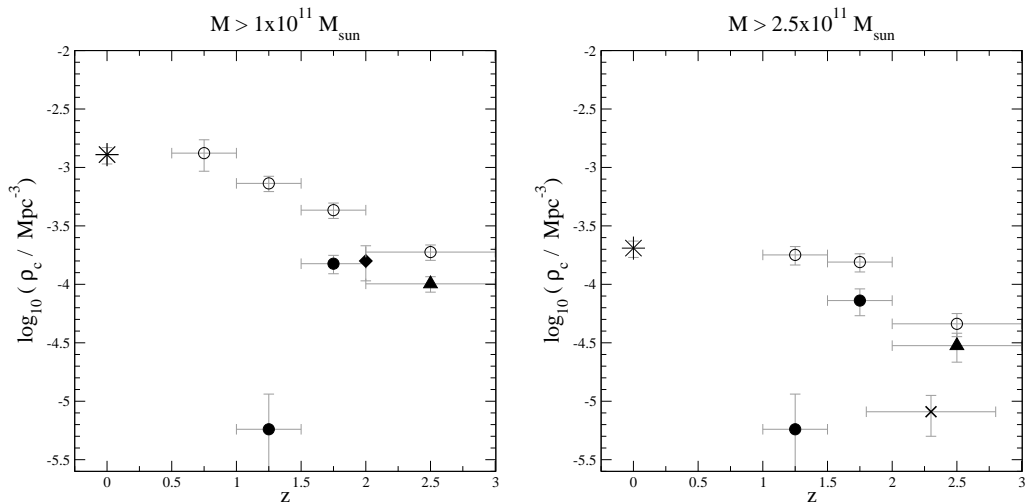}
\caption{The evolution of the comoving number density of $K_s<21.5$ (Vega)  galaxies with redshift (empty circles), in comparison to that of ULIRGs with the same mass cut (filled circles). Filled triangles represent lower limits on the ULIRG densities. The diamond-like symbol in the left panel shows the density of ULIRGs estimated by Daddi et al.~(2005). The cross-like symbol in the right panel corresponds to the density of radio-detected submillimetre galaxies with $S_\nu > 5 \, \rm mJy$ (Chapman et al.~2003). Figure taken from Caputi et al.~(2006b).}
\label{fig3}
\end{figure} 

Caputi et al. (2006b) studied the role of the LIRG and ULIRG phases in the evolution of massive $K_s$-band galaxies, using the same $24 \, \rm \mu m$ sample in the GOODS/CDFS as analyzed in Section 2. 
They found that LIRGs and ULIRGs only constitute a fraction of the massive ($M>10^{11}\, M_\odot$) galaxies present at different redshifts, but this fraction becomes very important ($>50\%$) at $z>2$.

Certainly, from Figure~\ref{fig3} we can see that ULIRGs trace a substantial fraction of massive galaxies at high redshifts. The density of ULIRGs sharply decreases below $z \approx 1.5$, but LIRGs still constitute $\sim 30\%$ of the galaxies with $M>10^{11}\, M_\odot$.

\section{The contribution of $24 \, \rm \mu m$ galaxies to the far-IR background}

In the previous section we have explained the importance of 24~$\mu$m galaxies within the population of massive galaxies. However, one could wonder whether  galaxies selected at 24~$\mu$m are actually  representative of the galaxy populations selected at other IR wavelengths.

To explore this issue, Dole et al.~(2006) studied the contribution of  {\em Spitzer/ MIPS} 24~$\mu$m galaxies to the Far-Infrared (FIR) Background at 70 and 160~$\mu$m, using stacking analysis over an area of 0.85~deg$^2$. This technique consist  in studying the integrated light at 70 and 160 of all the resolved 24~$\mu$m sources. Resolved 24~$\mu$m sources make up $\sim 80\%$ of the 24~$\mu$m background (Papovich et al.~2004; Dole et al.~2006). However, due to confusion, resolved 70 and 160~$\mu$m sources only can explain a minor fraction of the respective 70 and 160~$\mu$m backgrounds. Dole et al.~(2006) showed  that the stacking analysis allows to gain an order of magnitude below the confusion level.  They determined that 24~$\mu$m sources account for 92 and 69\% of the 70 and 160~$\mu$m backgrounds, respectively. This is the first measurement of the contribution of  24~$\mu$m galaxies to the far-IR cosmic background.

\begin{figure}
\centering 
\includegraphics[width=14cm]{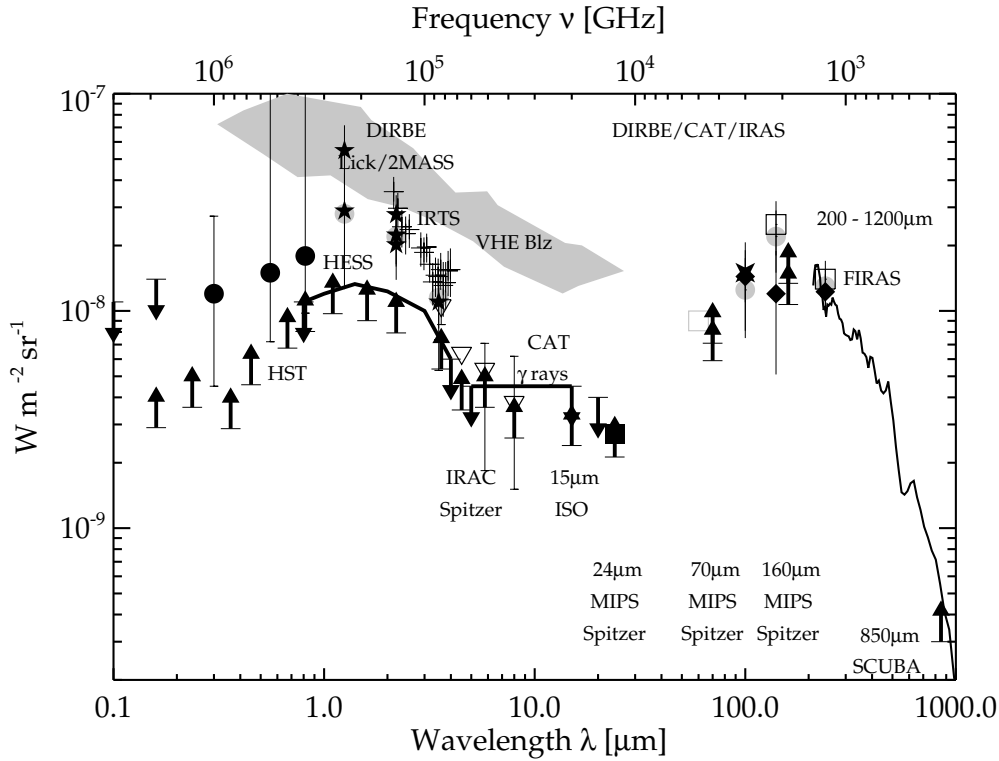}
\caption{ The EBL SED from 0.1~$\mu$m to 1~mm. The arrows at 24, 70 and 160~$\mu$m show the fraction of the CIB resolved through the stacking analysis of $S_\nu(24 \, \rm \mu m) > 60~\mu$Jy sources, as obtained by Dole et al.~(2006). See this paper for further references on this figure.}
\label{fig4}
\end{figure}

Figure~\ref{fig4} shows the SED of the Extragalactic Background Light (EBL). Surveys conducted with different UV/optical to submillimetre facilities have progressively allowed to put constraints on this SED at different wavelengths. The arrows at 24, 70 and 160~$\mu$m  indicate the lower limits on the IR background determined through the stacking analysis of 24~$\mu$m sources performed by Dole et al.~(2006). Unlike all previous determinations, the stacking analysis allowed to obtain a precise direct measurement of the EBL in this wavelength range.

\begin{figure}
\centering 
\includegraphics[width=8cm]{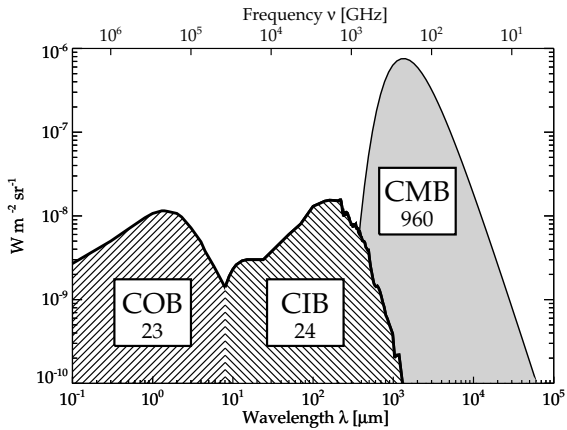}
\caption{Schematic SEDs of different extragalactic backgrounds. The numbers appearing inside the squares indicate the integrated intensities of the respective backgrounds, in units of nW~m$^{-2}$~sr$^{-1}$. The CMB contains $\sim 20$ times as much energy as the COB and CIB altogether. Figure taken from Dole et al.~(2006).}
\label{fig5}
\end{figure}

The determined intensity of the CIB is 24 nW~m$^{-2}$~sr$^{-1}$. This intensity is similar to that of the Cosmic Optical Background (COB).  Put another way, half of the energy associated to galaxy formation and evolution directly comes from starlight, while the other half is due to the light reprocessed by the dust. However, altogether, the energy budgets of the COB and the CIB  are equivalent to only 5\% of the energy contained in the Cosmic Microwave Background (CMB; Dole et al.~2006). This percentage represents the fraction of the energy which has been produced after recombination.  The comparison of intensities of the different background SEDs is illustrated in figure~\ref{fig5}.

\section{Conclusions}

{\em Spitzer} is making possible an unprecedented study of the mid-IR Universe and to set important constraints on the links between the mid-IR and other galaxy populations. On the one hand, we have shown that mid-IR sources constitute a significant fraction of the already-assembled massive galaxies at different redshifts. This implies that star-formation and accretion processes play a fundamental role in the evolution of massive galaxies through cosmic time. On the other hand, we have determined that those galaxies composing the 24~$\mu$m background are also responsible for most of the extragalactic energy produced at 70 and  160~$\mu$m. In conjunction, our results demonstrate  that the importance of studying mid-IR galaxies extends beyond the mid-IR domain  and this study is necessary to achieve a unified picture of galaxy populations.



\end{document}